\documentclass[preprint,showpacs,preprintnumbers]{revtex4}

\usepackage[dvips]{color}
\usepackage{graphicx}
\usepackage{epsfig}
\usepackage{dcolumn}
\usepackage{amssymb}
\usepackage{amsmath}
\usepackage[english]{babel}
\usepackage{mathrsfs}
\usepackage{bm}
\usepackage{lscape}

\newcommand{\ket}[1]{\left|#1\right\rangle}
\newcommand{\bra}[1]{\left\langle#1\right|}

\def\*#1{\mathbf{#1}}

\begin{document}

\title{
Controlling the dynamics of a coupled atom-cavity system by pure dephasing: 
                        basics and potential applications in nanophotonics
        }

\author{A. Auff\`eves$^{1}$}\email{alexia.auffeves@grenoble.cnrs.fr}
\author{D. Gerace$^{2}$}
\author{J.-M. G\'erard$^{3}$}%
\author{M. Fran\c ca Santos$^{4}$}%
\author{L. C. Andreani$^{2}$}
\author{J.-P. Poizat$^{1}$}%

\affiliation{$^{1}$ CEA/CNRS/UJF 
Joint team  ``Nanophysics and
semiconductors,''  Institut N\'eel-CNRS, \\ BP 166, 25, rue des Martyrs,
38042 Grenoble Cedex 9, France}

\affiliation{$^{2}$ Dipartimento di Fisica ``Alessandro Volta,'' and UdR CNISM,
                               Universit\`a di Pavia, via Bassi 6, 27100 Pavia, Italy}

\affiliation{$^{3}$ CEA/CNRS/UJF 
Joint team ``Nanophysics and
semiconductors,'' CEA/INAC/SP2M, \\ 17 rue des Martyrs, 38054 Grenoble, France}

\affiliation{$^{4}$ Departamento de F\'isica, Universidade Federal de Minas 
Gerais, Caixa Postal 702, 30123-970, Belo Horizonte, Brazil}


\date{\today}

\begin{abstract}
The influence of pure dephasing on the dynamics of the coupling between a 
two-level atom and a cavity mode is systematically addressed. 
We have derived an effective atom-cavity coupling rate
that  is shown to be a key parameter in the physics of the problem, 
allowing to generalize the known expression for the Purcell factor to 
the case of broad emitters, and to define strategies to optimize 
the performances of broad emitters-based single photon sources. 
Moreover, pure dephasing is shown to be able to restore lasing in 
presence of detuning, a further demonstration that decoherence 
can be seen as a fundamental resource in solid-state cavity quantum 
electrodynamics, offering appealing perspectives in the context of 
advanced nano-photonic devices. We propose experimental strategies 
to develop a new type of versatile device that can be operated either as a single photon 
source or as a laser, based on the control by decoherence of the coupling
between a single quantum dot and a solid-state cavity.
\end{abstract}

\maketitle

\section{Introduction}
Cavity Quantum Electrodynamics (CQED) aims at describing light-matter interaction when 
light and matter reduce to canonical systems, i.e. when light can be modeled by a 
single mode of the electromagnetic field and matter by a single two-level system. To achieve 
this situation, one should couple a given transition of the matter field to a monomode cavity. 
The losses and dephasing times of each system should happen on a time scale much longer than
the coupling timescale. Energy can thus be coherently exchanged between the atom and the cavity, allowing to implement fundamental tests of quantum mechanics, and opening the way to 
quantum information processing. On the way to the quantum regime, Purcell enhancement, 
which occurs when the
emitter lifetime is modified as a consequence of its resonant coupling to the cavity mode, 
is a well known milestone \cite{Purcell}. 
Historically, first CQED evidences were demonstrated with atoms coupled to microwave 
\cite{Goy} and optical \cite{Thompson} cavities, respectively. 
These systems are characterized by a very long radiation lifetime of the isolated emitter, 
which in the spectral domain corresponds to a very narrow dipole resonance coupled to a 
broad cavity. This picture has been the usual paradigm for CQED so far.

On the other hand, CQED experiments can now be performed with solid-state emitters 
(so called \textit{artificial atoms})  and cavities. The strong coupling regime has been 
reached for the excitonic transition of 
quantum dots \cite{stcouplqd} and nanocrystals \cite{LeThomas,Park_nl06} coupled to optical 
semiconductor cavities, as well as for superconducting qubits coupled to microwave 
cavities \cite{Wallraff}. In all of these systems, the cavity mode quality factor can be very large, 
while solid state emitters are intrinsically coupled to the matrix they are embedded in. 
In fact, decoherence and phase relaxation unavoidably broaden any transition between the 
discrete states of such artificial atoms. 
These new conditions open an unexplored regime for CQED so far,  where the emitter's linewidth 
can be of the same order of magnitude, or even broader than the cavity mode one.
Different mechanisms contribute to the decoherence-induced broadening 
of artificial atoms, among which phonon-assisted mechanisms \cite{Borri}, 
or spectral diffusion \cite{Cassabois}. 
If spectral diffusion happens on a timescale much shorter than the typical spontaneous 
emission timescale, it can safely be modeled by a simple pure dephasing channel in the 
master equation describing the dynamics of the system. 
Because of its simplicity, the scheme of a two-level system undergoing pure 
dephasing can be regarded as an appealing tool to explore this new regime of CQED, 
as well as a useful effective model to describe solid-state emitters \cite{Naesby, Alexia}. 

Such a model has mostly been used to study the spectral properties of the light 
emitted by the atom-cavity system so far, especially with the aim of  describing 
microphotoluminescence experiments performed on quantum dots coupled to semiconductor
cavities \cite{Laucht, LaussyI,Delvalle}.
In particular, pure dephasing has been identified as a potential mechanism for the so-called 
cavity feeding process \cite{Naesby,Alexia}, namely the emission of photons at the cavity frequency 
that shows up even if the emitter is coupled to a detuned cavity mode 
\cite{Hennessy,Press,Kaniber,Laucht,Pascale,Winger}.  
This picture for artificial atoms in the solid state has shed new light on the effects of decoherence, 
which can be considered as a supplementary degree of freedom as compared to isolated 
atoms, offering appealing perspectives to achieve advanced nanophotonic devices controlled 
by pure dephasing \cite{Alexia}. 

Nevertheless, little attention has been paid to the dynamics of an emitter subject to pure 
dephasing and coupled to a cavity up to now. 
In this work, we consider the temporal evolution of the emitter's and 
cavity mode populations, and show that the system in the incoherent regime can be 
described by a classical (rate-equation) model with an effective atom-cavity coupling rate. 
Such a coupling rate is a key parameter: it allows to define a generalized Purcell factor, 
and beyond, to revisit the notions of good and bad cavity regimes, respectively.
We show that pure dephasing can increase the effective atom-cavity coupling, thus enhancing 
the effective Purcell factor of the system. 
Switching to the non-linear regime, we also show that the lasing properties of a single emitter can be evidenced by looking at
the statistical properties of the emitted light, from which we characterize the conditions 
for the lasing onset to be achieved. 
In particular, we define parameters for which pure dephasing can even induce 
lasing, thus showing that decoherence can turn out to be an extremely positive resource 
in the framework of prospective solid-state nanophotonic devices. 
Within these framework, quantum dots appear as promising candidates, as the 
experimental strategies to tune their environment, thus the pure dephasing rate that
controls their homogeneous linewidth, have already started to be developed. As 
a consequence, we have mostly used parameters consistent with state of the art experiments
using quantum dots coupled to optical semi-conducting cavities. On top of it, we propose to
implement a versatile device, that can be operated either as a single photon source 
or as a laser, based on the control of decoherence induced by the environment of the quantum dot.
We also propose an innovative strategy to control this environment.

The paper is organized as follows. We first introduce an effective atom-cavity coupling parameter, 
and use it to define the good and bad cavity regimes, when pure dephasing is properly taken 
into account. 
Focusing our attention onto the bad cavity regime, we derive a generalized Purcell factor, and 
analyze how to optimize the rate of emitted photons by a broad emitter-based single photon source. 
We then address the good cavity regime by revisiting the properties of the single emitter laser, 
paying attention to the influence of pure dephasing. We finally analyze the relevance of
the developed model to the case of quantum dots coupled to semiconductor photonic cavities.

\section{Effective atom-cavity coupling}

\begin{figure}[t]
\begin{center}
\includegraphics[width=9cm]{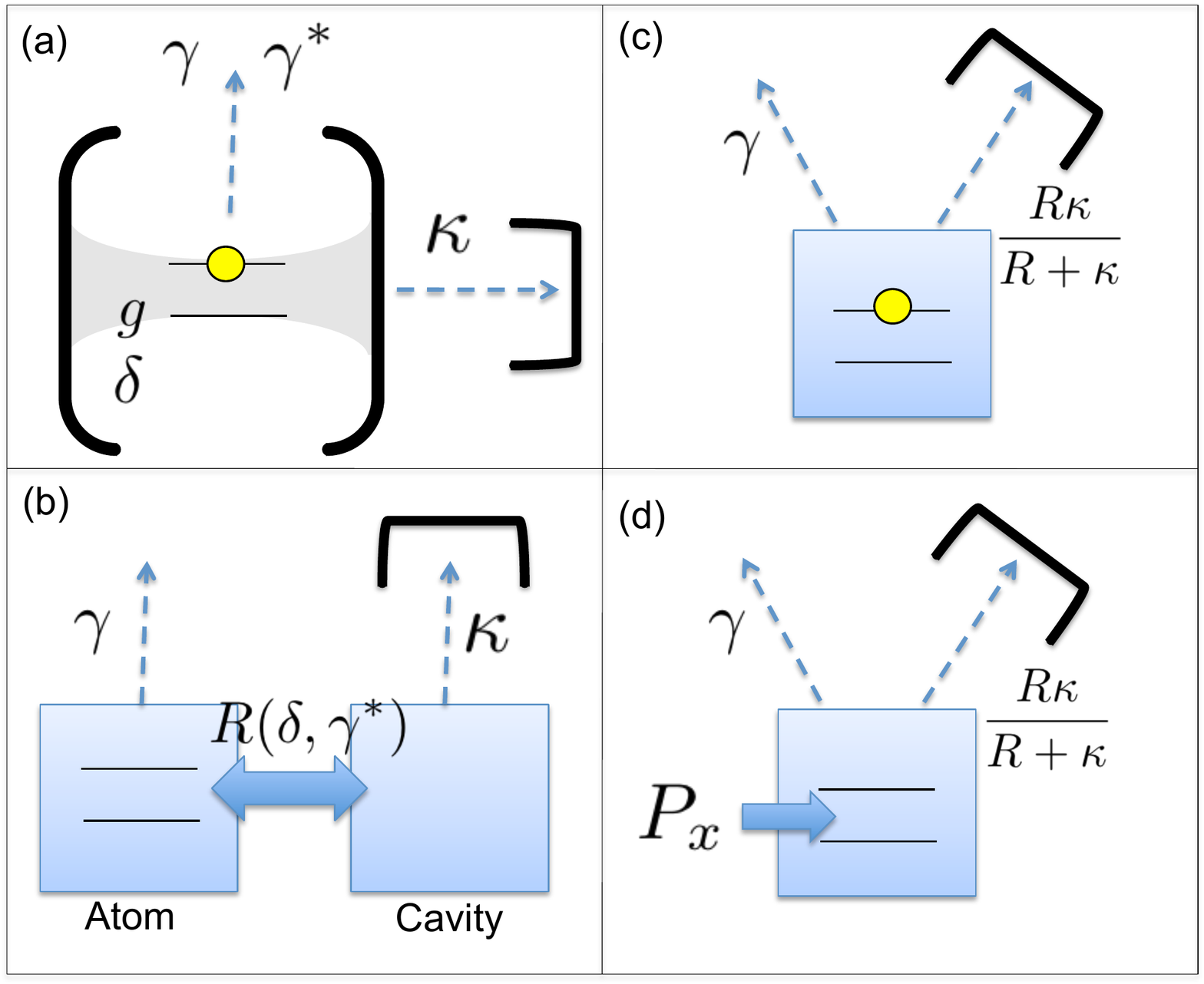}
\caption{(Color online).
(a) System under study: a two-level atom undergoing pure dephasing coupled to a cavity. 
(b) Equivalent classical system: two connected boxes exchanging a quantum of energy. 
(c) Equivalent system in the spontaneous emission regime, and  
(d) in the continuous pumping regime, respectively. } \label{fig:systeme}
\end{center}
\end{figure}

The system under study is represented in Fig.~\ref{fig:systeme}a. 
A two-level atom of frequency $\omega_x$  is initially pumped in its excited state. 
It is coupled to a single mode cavity of frequency $\omega_a$,  with a strength $g$. 
The detuning between the atom and the cavity is denoted $\delta=\omega_x-\omega_a$, 
the losses from the isolated atom and from the cavity mode are respectively 
$\gamma$ and $\kappa$. 
The atom undergoes pure dephasing with a rate $\gamma^{\ast}$.  
One first recalls some results related to the relaxation of the atom-cavity system, the 
atom having initially been excited \cite{Alexia}.
The full quantum evolution of this system is described by the master equation
\begin{equation}\label{equ:evol}
\dot{\rho}=i[\rho,\hat{H}]+{\cal L}_{damp}^{cav}+{\cal L}_{damp}^{at}+{\cal L}_{deph} \, ,
\end{equation}
where the total Hamiltonian of the system is ($\hbar=1$) 
\begin{equation}\label{eq:model}
\hat{H}= \omega_x \hat{\sigma}_+ \hat{\sigma}_- + \omega_{a} \hat{a}^{\dagger} \hat{a} 
                +  i g(\hat{a}^{\dagger} \hat{\sigma}_-  - \hat{\sigma}_+ \hat{a} ) \, .
\end{equation}
Here, $\hat{a}$ ($\hat{a}^{\dagger}$) is the annihilation (creation) operator for a photon in 
the cavity mode, while
$\hat{\sigma}_-$ ($\hat{\sigma}_+$) is the lowering (rising) operator for the atom.
The damping part for both the atom and the cavity mode can be described by operators 
in the Lindblad form within the master equation, written as
\begin{eqnarray}\label{damp}
 {\cal L}_{damp}^{(cav)} &=& 
 \frac{\kappa}{2} (2 \hat{a}\rho \hat{a}^\dagger - \hat{a}^\dagger \hat{a} \rho - 
 \rho \hat{a}^\dagger \hat{a}) \nonumber \\
 {\cal L}_{damp}^{(at)} &=& 
 \frac{\gamma} {2} (2 \hat{\sigma}_- \rho \hat{\sigma}_+ -\hat{\sigma}_+ \hat{\sigma}_- \rho - 
 \rho\hat{\sigma}_+ \hat{\sigma}_-) \, .
\end{eqnarray}
We notice that in most of the experimental situations currently accessible, the spontaneous 
emission rate $\gamma$ is much lower than the other typical rates, particularly the cavity 
damping rate $\kappa$. In the following, we will keep the spontaneous emission rate in the formulas
but shall neglect this parameter in comparison with the cavity decay rate as far as possible. The pure dephasing channel is described by the following Lindblad operator:
\begin{equation}\label{deph}
{\cal L}_{deph} =   \frac{\gamma^{\ast}} {4} ( \hat{\sigma}_z \rho \hat{\sigma}_z - \rho) \, .
\end{equation}

From the master equation, the evolution of the populations and coherences is described by 
the following equations of motion:
\begin{eqnarray}\label{equ:populations}
\frac{d \langle \hat{a}^\dagger  \hat{a} \rangle }{dt} &=& 
-\kappa \langle  \hat{a}^\dagger  \hat{a}  \rangle + 
g \langle \hat{\sigma}_+  \hat{a} \rangle + g \langle  \hat{a}^\dagger \hat{\sigma}_- \rangle  \nonumber \\
\frac{ d \langle \hat{\sigma}_+\hat{\sigma}_- \rangle}{dt} &=&
-\gamma \langle \hat{\sigma}_+\hat{\sigma}_- \rangle - g \langle \hat{\sigma}_+  \hat{a} \rangle 
- g \langle  \hat{a}^\dagger \hat{\sigma}_- \rangle    \nonumber \\
\frac{ d \langle \hat{\sigma}_+ \hat{a} \rangle} {dt}  &=&
i\delta \langle \hat{\sigma}_+ \hat{a} \rangle -\frac{\gamma+\gamma^{\ast}+\kappa}{2} \langle
\hat{\sigma}_+ \hat{a} \rangle \nonumber \\
&& + g (\langle \hat{\sigma}_+ \hat{\sigma}_- \rangle - \langle
\hat{a}^\dagger  \hat{a} \rangle ) \,  .
\end{eqnarray}
The coherent or quantum regime is characterized by the reversible exchange of a quantum 
of energy between the atom and the cavity mode. 
At resonance, this so-called vacuum Rabi oscillation shows up if the vacuum Rabi splitting 
$2g$ essentially overcomes the effective dephasing rate
$\gamma+\gamma^{\ast}+\kappa$. 
The opposite case corresponds to a regime where the energy is irreversibly spread between 
the atom, the cavity and the environment. 
In this incoherent, classical regime, the crossed terms $\langle  \hat{a} ^\dagger \hat{\sigma} \rangle$ and 
$\langle \hat{\sigma}_+  \hat{a} \rangle$, which are responsible for the Rabi oscillation, 
can be adiabatically eliminated. 
Out-of-resonance, the adiabatic elimination is valid for $\delta>g$.

At this step it is worth pointing out the difference between the coherent regime and the so-called 
strong coupling regime.
Strong coupling is reached when the spontaneous emission spectrum of the system consists 
in two peaks of distinct frequencies \cite{Claudio}, which corresponds, taking into account pure 
dephasing, to $2g>|\gamma+\gamma^{\ast}-\kappa|$. If this condition is fulfilled, the width of each 
peak equals $(\kappa+\gamma+\gamma^{\ast})/2$. Coherent coupling is reached when the doublet 
is resolved, which is the spectral counterpart of the vacuum Rabi oscillation. 
In the standard CQED picture where the atomic width $\gamma^{\ast}+\gamma$ is negligible, the 
strong coupling regime perfectly matches the coherent regime.
On the contrary, when $\gamma^{\ast}$ is of the same order of magnitude as $\kappa$, strong 
coupling is a necessary but not sufficient condition to observe a coherent exchange of energy 
between the atom and the cavity mode. 
Coherent regime thus appears as more demanding than strong coupling.

In the incoherent regime, the adiabatic elimination leads to the set of coupled
dynamical equations: 
\begin{eqnarray}\label{equ:adiab}
\frac{d \langle \hat{a}^\dagger \hat{a} \rangle }{dt}&=&
-(\kappa + R) \langle \hat{a} ^\dagger \hat{a} \rangle + 
R \langle \hat{\sigma}_+ \hat{\sigma}_- \rangle \nonumber \\
\frac{ d \langle \hat{\sigma}_+\hat{\sigma}_- \rangle} {dt} &=& 
-(\gamma + R) \langle\hat{\sigma}_+ \hat{\sigma}_- \rangle + 
R \langle \hat{a} ^\dagger \hat{a}  \rangle \, ,
\end{eqnarray}
where  we have introduced the quantity 
\begin{equation}\label{eq:rcoupl}
R=\frac{4g^2}{\kappa+\gamma+\gamma^{\ast}}
\frac{1}{1+\left( \frac{2\delta}{\kappa+\gamma+\gamma^{\ast}}\right)^2} \, \, .
\end{equation}
As it was underlined in Ref.~\cite{Alexia}, the quantity $R$ can be seen as an effective 
coupling rate between the atom and the cavity mode, i.e.
the system is formally equivalent to two coupled boxes (as represented in Fig.~\ref{fig:systeme}b). 
The ``atomic'' box is initially charged with a quantum of energy that can escape in the environment 
at rate $\gamma$, or in the ``cavity'' box at rate $R$. In the same way, the cavity box can lose 
its excitation with a rate $\kappa$, or give it back to the atomic box with a probability per unit 
time $R$. 
The parameter $R$ is also involved in the efficiency of the corresponding single 
photon source, which reads
\begin{equation}\label{eq:beta}
\beta=\frac{R\kappa/(R+\kappa)}{\gamma+R\kappa/(R+\kappa)} \, .
\end{equation}

Seen from the atom point of view, the cavity mode appears as a further loss channel, 
whose effective rate is $R\kappa/(R+\kappa)$ (see Fig.\ref{fig:systeme}c). 
This result could have been straightforwardly derived from the classical picture. 
Such an expression for $\beta$ is valid in any regime, even out of the incoherent regime, 
the only one in which the adiabatic elimination is supposed to be valid. 
Thus, the effective coupling rate $R$ appears as a key parameter,  allowing us to revisit 
the notions of good and bad cavity regimes, respectively. 
The bad cavity regime is achieved when $\kappa > R$, namely when the cavity 
damping time is shorter than the typical atom-cavity coupling time. In this case, 
the quantum exits the cavity mode as soon as it is released from the emitter: the cavity 
behaves as a supplementary loss channel. 
This is the usual regime for single photon sources and it is studied in the next two sections. 
In the good cavity regime, which is achieved when $R>\kappa$, the quantum of energy
is emitted by the atom and can stay in the cavity mode before being lost in the environment. 
We stress here that the good cavity regime, which at resonance is achieved when 
$2g>\sqrt{\kappa(\kappa+\gamma+\gamma^{\ast})}$, is more demanding than the strong coupling 
regime if $\gamma^{\ast}$ becomes non negligible (whereas again, the two regimes co\"incide 
in the standard CQED picture).
On the contrary, the good cavity regime does not necessarily imply a coherent energy 
exchange between  the atom and the cavity (which on resonance requires 
$2g>\kappa+\gamma+\gamma^{\ast}$), as it would be the case in usual CQED experiments 
performed with atoms. 
On the contrary, pure dephasing opens a new regime where the quantum 
of energy can stay in the cavity mode without being reabsorbed by the atom. 
As it will be studied in section V, the good cavity regime is a necessary condition to 
implement single emitter lasers. 

\section{Generalized Purcell factor}

In this section, we focus on the bad cavity regime and  show that the effective coupling 
$R$ allows to define a generalized Purcell factor. 
By definition, in this regime the cavity behaves like a source of losses, and the atom-cavity 
coupling is incoherent. As a consequence, the parameter $R$ has the dynamical meaning 
of an effective spontaneous emission rate. If $R\ll \kappa$ (which corresponds to the so called 
Purcell regime), one can easily extract the atomic relaxation rate from the set of 
Eqs.~(\ref{equ:adiab}), which is $ \gamma + R$. As expected, ``switching on''  
the cavity mode corresponds to creating an additional relaxation channel for the atom, 
whose typical rate is $R$. One can thus define a generalized Purcell factor $F^{\ast}=R/\gamma$, 
quantifying the enhancement of spontaneous emission rate that simultaneously takes into 
account the influence of pure dephasing. This factor can be expressed as
\begin{equation}
F^{\ast}=\frac{4g^2}{\gamma(\kappa+\gamma+\gamma^{\ast})}
\frac{1}{1+\left(\frac{2\delta}{\kappa+\gamma+\gamma^{\ast}} \right)^2} \, .
\end{equation}
We can notice that one recovers the usual expression for the Purcell factor, 
$F=4g^2/\kappa\gamma$ \cite{jmg,jmg2,Serge}, for  $\gamma^{\ast}=0$. 
With respect to the standard expression, $F^{\ast}$ is obtained by replacing the cavity 
mode linewidth, $\kappa$, with the sum of  $\kappa$ and the total emitter's linewidth, 
$\gamma+\gamma^{\ast}$.  This essentially reduces to replacing the bare cavity mode 
Q-factor,  $Q_{cav}=\omega_{cav} / \kappa$, which usually appears in the standard 
Purcell expression,  with an effective quality factor, $Q_{eff}$, depending also on the 
emitter's quality factor, $Q_{em} = \omega_0 /(\gamma+\gamma^{\ast})$, as
\begin{equation}
\frac{1}{Q_{eff}} = \frac{1}{Q_{cav}}+\frac{1}{Q_{em}} \, . 
\end{equation}

\begin{figure}[t]
\begin{center}
\includegraphics[width=9.5cm]{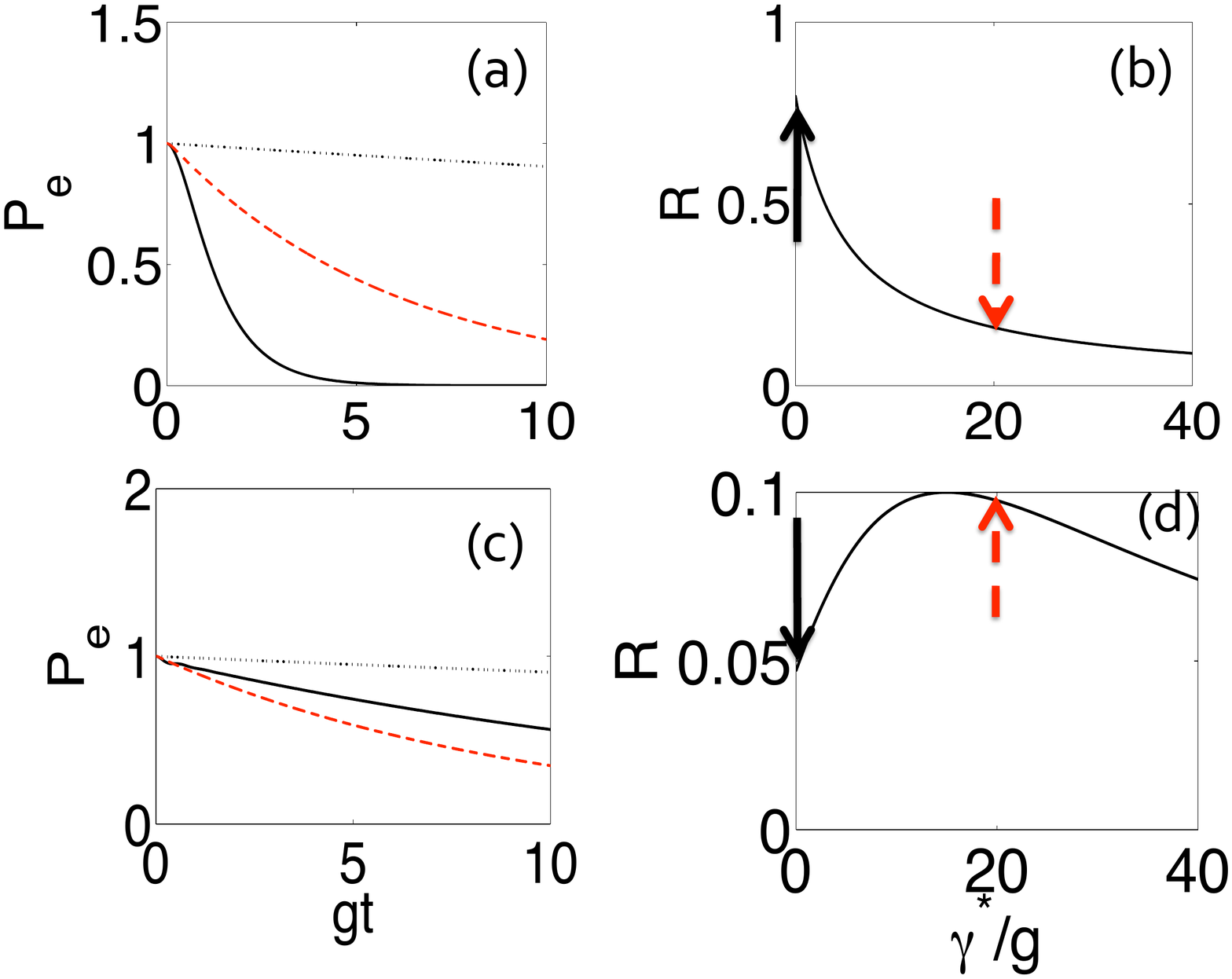}
\caption{ (Color online).
Evolution of the atomic population as a function of time for resonant (a) and
(b) detuned (with $\delta=10 g$) case. 
Dotted line: the atom and the cavity are not coupled.
Solid black curve: $\gamma^{\ast}=0$. 
Dashed red curve: $\gamma^{\ast}=20 g$. 
Corresponding behavior of the effective atom cavity coupling, $R$, 
as a function of pure dephasing is shown for both (b) resonant and (d) detuned cases. 
The black arrow indicates the case where $\gamma^{\ast}=0$ [corresponding to the red curves 
in (a) and (c)], while the red dashed arrow is for the case $\gamma^{\ast}=20 g$.
[corresponding to the red curves in (a) and (c)]. 
The other parameters of the model for these calculations are: 
$\gamma=0.01g$, $\kappa=5g$. } \label{fig:Purcell}
\end{center}
\end{figure}

The existence of a generalized Purcell factor had already been
heuristically derived in \cite{jmg,Yokohama} and finds here a demonstration in the case 
of a single emitter homogeneously broadened.
Note that this effective quality factor gives a symmetrical role to the emitter and to the cavity 
mode as long as one does not exit the bad cavity regime. 
In experiments exploiting CQED effects, increasing the Purcell factor is an important goal, 
giving rise to a quest to increase the quality factor of the cavity mode. According to the 
generalized expression we have derived, it makes sense to search for the highest possible 
$Q_{cav}$ (even when it overcomes $Q_{em}$) compatible with the bad cavity regime.  
In particular, by making $Q_{cav}$ bigger than $Q_{em}$, one can double the effective Purcell 
factor of the system in the case where the atom is resonant with the cavity mode. 
On the contrary, increasing pure dephasing $\gamma^{\ast}$ leads to a decrease of the emitter's 
quality factor $Q_{em}$, thus to a decrease of the effective quality factor $Q_{eff}$. 
Consequently, the spontaneous emission rate is reduced as it appears in Fig.~\ref{fig:Purcell}a,
where we have represented the relaxation of an initially excited atom in the resonant case for 
different values of the pure dephasing rate. 

The influence of pure dephasing is dramatically different if the atom and the cavity are detuned, 
as already pointed out in  Ref.~\cite{Noda}. In this case indeed, it makes sense that a decrease 
of $Q_{em}$, thus of $Q_{eff}$, induced by pure dephasing leads to the enhancement of 
the spontaneous emission rate as it clearly appears in Fig.~\ref{fig:Purcell}b. To quantify the 
maximal enhancement we can get, it is worth noticing that the parameter $R$ is maximized 
when $\kappa+\gamma+\gamma^{\ast}=2\delta$, allowing to reach an optimal value 
$R_{max}=2g^2/\delta$ and a maximal effective Purcell factor $F^{\ast}_{max} = 2g^2/\delta\gamma$.
Higher values of pure dephasing rate lead to a decrease of the effective atom-cavity coupling, 
and consequently of the spontaneous emission rate.
By playing on pure dephasing, one can significantly improve the effective Purcell factor by a 
factor $F^{\ast}_{max} /F \sim \delta / \kappa$. 

Note that we have used state of the art parameters of quantum dots coupled to optical 
semi-conducting cavities, where coupling strengths like $g=50$ $\mu$eV and cavity 
linewidths $\kappa = 250$  $\mu$eV (corresponding to quality  factors of $Q_{cav}\sim 5000$) 
are commonly reached.  Tuning of the pure dephasing rate can be achieved using temperature 
or pump power  as it clearly appears in the studies described in \cite{Favero}. 
The observed behavior was attributed to the fluctuation in the occupancy of electron traps in the 
QD neighbourhood \cite{Cassabois}. 
In particular, these experiments show that the linewidth of the QD exciton can be tailored 
almost at will over a large range of experimental values  (1-500 $\mu$eV). 
Nevertheless, non-resonant pumping was used in these studies, thus allowing the trapping of more 
than a single exciton in the quantum dot and effectively hindering the validity of the two-level 
atom model  used in the present paper. 
To preserve the validity of our approach, quasi-resonant pumping 
(either in the p-shell, or using a resonant mechanism assisted by the creation of a phonon)
or direct resonant pumping should be used \cite{Ates, Englund}, where the 
pump power is expected to have no influence on the density of carriers around the quantum 
dot and thus has no influence on the pure dephasing rate.
Moreover, as the temperature also acts on the QD-cavity detuning, it is necessary to use 
another parameter to adjust it,  for instance the electric-field \cite{Laucht} or a tunable 
microcavity \cite{Kaniber}. 
Another promising approach could consist in optically pumping a quantum well positioned in 
the vicinity of the quantum dot:  by controlling the carrier density in the quantum well, one could 
adjust the amount of decoherence induced by Coulomb interaction between the trapped exciton 
and the nearby quantum dot under study. The development of such a device would allow to 
explore the new regimes for CQED studied in the present paper and 
directly check the influence of the pure dephasing rate on the dynamics of the QD cavity coupling. 
Moreover it would realize a novel type of nano-photonic device based on the exploitation of 
decoherence, showing that pure dephasing is a resource specific to solid-state emitters, as it will
be shown in the following examples. In particular, we propose strategies based on this control
to optimize the figures of merit of single photon sources and nanolasers.

\section{Broad emitter-based single photon sources}

A figure of merit usually considered for single photon sources is the efficiency of 
the device. Assuming that  the detector is 
perfectly geometrically coupled to the cavity channel of losses, this corresponds to 
the probability for the photons to be spontaneously emitted in the cavity mode, which
is usually denoted $\beta$.
The behavior of this parameter, also considering the effects of pure dephasing and 
atom-cavity detuning,  is studied in Ref.~\cite{Alexia},
However, the system was studied in the spontaneous emission picture, 
which does not model a usual photoluminescence experiment where the atom is 
typically pumped in continuous wave. In this context, an interesting figure of merit for a single 
photon source is the rate of photons that is emitted in the cavity loss channel, ${\cal N}$. 
To describe the pumping mechanism on the atom, one has to add to the master equation a 
Lindblad operator that is formally expressed as

\begin{equation}\label{equ:px}
{\cal L}_{pump}^{at}= 
\frac{P_x}{2} (2 \hat{\sigma}_+ \rho  \hat{\sigma}_-  -  \hat{\sigma}_- \hat{\sigma}_+ \rho - 
\rho  \hat{\sigma}_- \hat{\sigma}_+ ) \, .
\end{equation}
If we restrict our analysis to the bad cavity regime, we can safely assume that the 
dynamics is restricted to the subspace spanned by the three states 
$\{\ket{g,0}$, $\ket{g,1}$, $\ket{e,0}\}$, irrespective
of the pumping rate $P_x$. This allows us to obtain steady state solutions for the atomic 
($n_x= \langle \hat{\sigma}_+ \hat{\sigma}_- \rangle_{ss}$) and the cavity mode 
($n_a= \langle \hat{a}^\dagger \hat{a}  \rangle_{ss}$) populations, respectively.
We obtain
\begin{eqnarray}\label{equ:ssqr}
n_x &=& \frac{P_{x}}{P_x+(\gamma+\frac{\kappa \tilde{R}}{\kappa + \tilde{R}})}  \\
n_a &=& \frac{\tilde{R}}{\kappa + \tilde{R}} n_x  \, \, ,
\end{eqnarray}
where we have introduced the effective rates
\begin{eqnarray}
\frac{\Gamma}{2}&=& \frac{P_x+\gamma+\gamma^{\ast}+\kappa}{2}   \label{equ:Gam}  \\
\tilde{R}&=& \frac{4g^2}{\Gamma}\frac{1}{1+(2 \delta/\Gamma)^2}  \label{equ:R}     \, \, . 
\end{eqnarray}
In the case where $P_x \ll \gamma+\gamma^{\ast}+\kappa$, one obtains for $\tilde{R}$ 
the expression of the effective coupling $R$ defined in Eq.~(\ref{eq:rcoupl}), 
making transparent the expression of the atomic population $n_x$. 
It corresponds indeed to the incoherent pumping 
of a two-level system connected to two different loss channels, 
the first being due to the continuum of leaky photonic modes
with a rate $\gamma$, and the second to the coupling to the cavity mode 
as it appears schematically in Fig.~\ref{fig:systeme}d. The effective loss rate of this 
second channel is, again, $\kappa \tilde{R}/(\kappa +\tilde{R})$, as already showed in 
the spontaneous emission picture. 
The latter considerations reinforce the generality of the physical interpretation 
for the parameter $R$. In the following, we identify the two coupling strengths and use a 
unified notation for it, $R$.

Finally, the rate of photons emitted in the bad cavity regime can be explicitly given as
\begin{equation}
{\cal N}=\kappa n_a=\frac{\kappa R}{\kappa + R}  
\frac{P_x}{\left(P_x+\gamma+\frac{\kappa R}{\kappa +R}\right) } \, \, .
\end{equation}
In the limiting case of very low pumping rate, i.e. for 
$P_x \ll \gamma + \kappa R/(\kappa + R)$, the rate of photons exiting the cavity 
can be expressed as ${\cal N}= \beta P_x$, where $\beta$ is the efficiency of the single 
photon source given in eq.(\ref{eq:beta}), showing that the low incoherent pumping 
scheme can safely be modeled as a series of spontaneous emission events with a 
rate $P_x$.
This conclusion completely restores the continuity between the spontaneous emission 
picture and the incoherent pumping one, and it justifies to optimize the efficiency 
$\beta$ when the device is operated below saturation.  
On increasing pump power, the rate of emitted photons saturates
to the value ${\cal N}_{sat}=\kappa R/(\kappa + R)$, which in the bad cavity regime 
reduces to ${\cal N}_{sat}= R$. 
In order to maximize the rate of emitted photons, one has to maximize the parameter
$R$, or equivalently $F^{\ast}$, either by increasing $Q_{eff}$ in the resonant case, or by 
lowering it, by playing on pure dephasing, in the detuned case. Note that the behavior 
of the parameter ${\cal N}$ with respect to the atom-cavity detuning, 
$\delta$, is dramatically different depending on the atom being saturated or not. 
If the pump power is low, then ${\cal N}$ evolves as $\beta(\delta)$, whereas if the pump 
power is high, ${\cal N}$ evolves like the atom-cavity coupling $R(\delta)$. 
Such a change of behavior turns out to be a fruitful method, e.g., to measure the Purcell 
factor of a single quantum dot coupled to a semiconductor microcavity, as 
it was evidenced in Ref.~\cite{Mathieu}.

\section{Single two-level emitter laser}

In the previous Section we have evidenced that in the bad cavity regime, 
the rate ${\cal N}$ of photons emitted by the cavity first evolves linearly with 
respect to pump power (same as the atomic population), before  reaching an upper 
bound imposed by the spontaneous emission rate $R$ (while the atomic population 
gets totally inverted). This limit is due to the saturation of the two-level emitter. 
A way to overcome it is to reach the stimulated emission regime, where the atom-cavity
coupling scales like the number of photons in the cavity. In this case, the system is 
operated as a single emitter laser. This ideal device where the gain medium is 
quantified at the single emitter level, is of tremendous conceptual interest, and has 
motivated many fundamental studies since \cite{Mu}. 
The primary interest of the lasing regime is that photons are mostly 
funneled into the cavity mode, allowing to achieve the highly efficient 
conversion of the incoherent power carried by the pump into single mode light. 
Second, as the emitter's cycling rate is enhanced, it can be pumped at a much 
higher rate before saturation is reached. In this Section we revisit the single atom 
laser topics, building on the notion of good cavity regime. 
We examine to which extent pure dephasing can be a resource in the frame of 
solid-state lasers, and relate this study  to recent experimental demonstrations of 
single quantum dot lasers.

\subsection{Why the good cavity regime is mandatory}
Here we show that the good cavity regime is a necessary condition to reach stimulated 
emission.  A heuristic demonstration has been developed in \cite{Pelton,jmg} and is based
on the search of proper conditions to reach a steady state cavity population $n_a \sim 1$: 
in this view, the production rate of photons in the cavity, which is at most $\beta P_x$, should 
overcome the cavity mode dissipation rate $\kappa$.
As the cycling rate $P_x$ is limited by the typical spontaneous emission rate $R$, 
a necessary condition can be formulated as $R>\kappa$, which is the condition for
good cavity regime. 

This reasoning is confirmed by analyzing the rate equations for the incoherently 
pumped atom, which can be derived from Eqs. (\ref{equ:populations}) and  (\ref{equ:px}), 
respectively. One finds
\begin{eqnarray}
\frac{\mathrm{d}\langle \hat{\sigma}_z \rangle}{\mathrm{d}t}
&=& -(R+\gamma)(\langle 1 + \hat{\sigma}_z \rangle)+P_x(1- \langle \hat{\sigma}_z \rangle)
-2R\langle \hat{a}^\dagger \hat{a}\rangle\langle \hat{\sigma}_z \rangle     \label{equ:SAL1}
\\
\frac{\mathrm{d}  \langle \hat{a}^\dagger \hat{a}\rangle}{\mathrm{d}t}
&=& \frac{R}{2}( 1 + \langle \hat{\sigma}_z \rangle)+
R \langle \hat{a}^\dagger \hat{a}\rangle \langle \hat{\sigma}_z \rangle
-\kappa \langle \hat{a}^\dagger \hat{a}\rangle   \label{equ:SAL2}  \, \, ,
\end{eqnarray} 
where $\hat{\sigma}_z=\ket{e}\bra{e}-\ket{g}\bra{g}$ stands for the population inversion,  
while the parameter $R$ is still the effective atom-cavity coupling defined in Eq. (\ref{equ:R}).
As it appears in Eqs. (\ref{equ:SAL1}-\ref{equ:SAL2}), 
the evolution of the population inversion depends on three
terms respectively due to spontaneous emission, pumping, and stimulated emission, whereas 
the evolution of the cavity mode involves spontaneous emission, stimulated emission and 
cavity losses. In steady state, the atomic population inversion 
${\cal I}=\langle \hat{\sigma}_z \rangle_{ss}$ and the cavity population 
$n_a= \langle \hat{a}^\dagger \hat{a}\rangle_{ss}$ are coupled in  the following 
way: 
\begin{eqnarray}
{\cal I} &=&   \frac{P_x-(R+\gamma)} {R(1+2n_a)+\gamma+P_x}  \\
n_a     &=&   \frac{1}{2} \frac{R({\cal I}+1)}{\kappa-R{\cal I}}  \, \, .
\end{eqnarray}

\begin{figure}[t]
\begin{center}
\includegraphics[width=8cm]{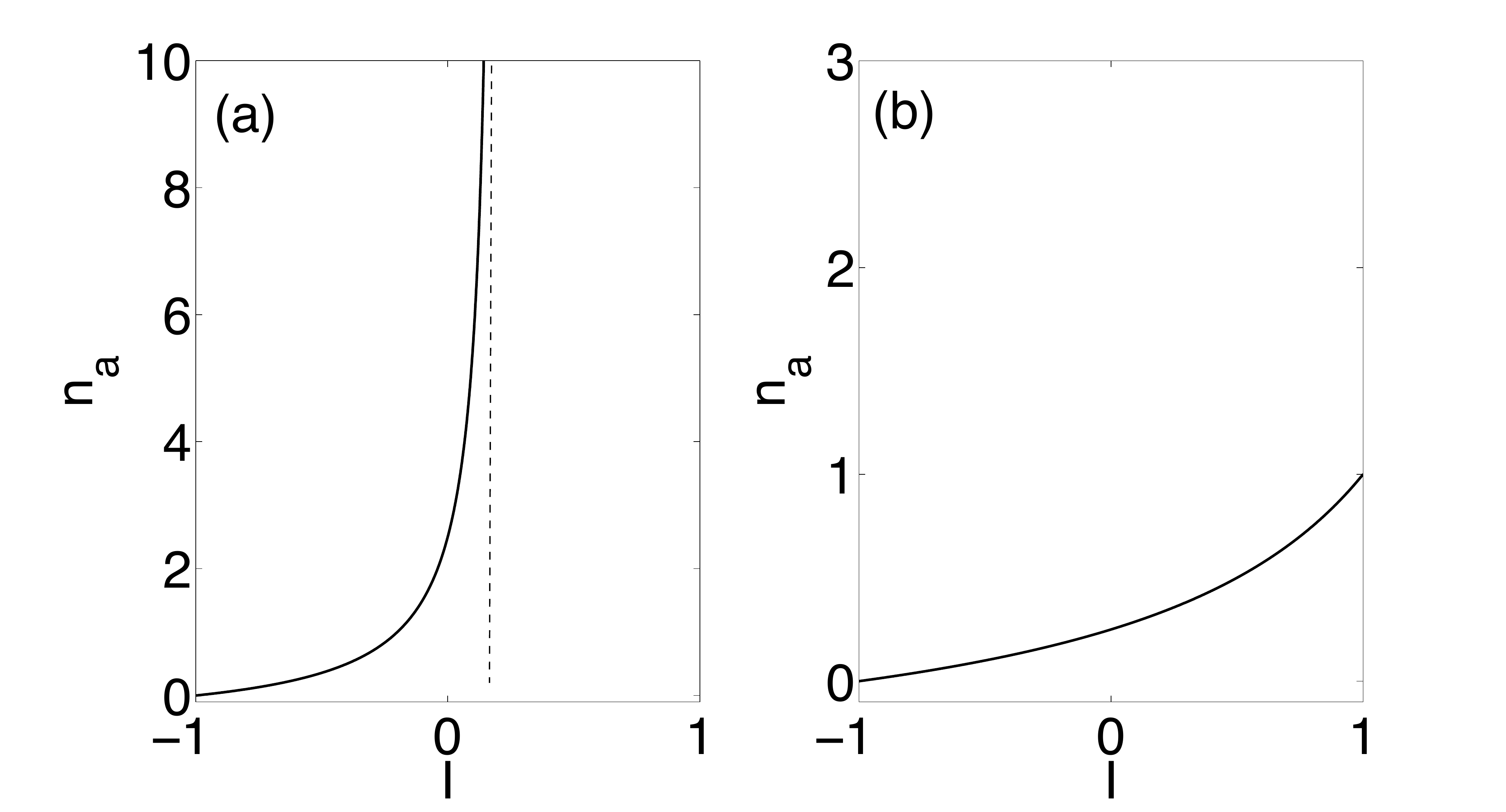}
\caption{Evolution of the steady state cavity population 
$n_a$ with respect to population inversion ${\cal I}$
We took $\kappa=0.2g$. (a) $R=g$ (good cavity regime); (b) $R=0.1g$ 
(bad cavity regime).} \label{fig:rate}
\end{center}
\end{figure}

We have represented in Fig.~\ref{fig:rate} the behavior of the steady state cavity population, 
$n_a$, with respect to  the atomic inversion, ${\cal I}$, by keeping the parameter $R$ constant,  
both in the good and bad cavity regimes. 
When the population is not inverted (${\cal I} \rightarrow 0$), which happens for low pumping 
rates, $n_a$ evolves linearly with respect to ${\cal I}$: this is the spontaneous emission regime. 
On the contrary, if stimulated emission can be reached, cavity population evolves linearly 
with the pump rate, while the atomic population inversion remains clamped to a value where the 
optical gain compensates for the losses. 
This behavior is characteristics of lasing, whatever the type of device (conventional lasers, 
high $\beta$ lasers, nanolasers). 
Namely, in this highly non-linear regime, the cavity population diverges with respect to the 
atomic one. As it appears in the figure, this can only happen in the good cavity regime, 
confirming the prediction above. 

Two strategies can be adopted to enter the good cavity regime: 
decreasing the cavity losses, i.e.  $\kappa$, or increasing the effective atom-cavity 
coupling, $R$. 
The first approach has been explored to study the potential of a quantum dot  coupled to a 
high Q microsphere to show lasing \cite{Pelton}. 
The other approach is more promising, as it allows to simultaneously increase the 
fraction $\beta$ of photons spontaneously emitted in the cavity mode. 
In the limit where $\beta\to 1$, the device has a perfect quantum efficiency, even before 
stimulated emission is reached. In such kind of devices, no kink can be observed in the 
input-output curve (namely, the rate of  emitted photons with respect to pump power), 
thus justifying the denomination of  ``thresholdless laser''  \cite{Rice,Bjork}. 
In the following we restrict the study to the case of a high $\beta$ single atom laser, 
and define signatures of the lasing regime. 

\subsection{Single two-level emitter lasing criteria}

We have studied the evolution of three main properties of the single emitter device 
with respect to pump power: 
the cavity population, $n_a$, the atomic population, $n_x$, and the 
auto-correlation function of the field at zero time delay, defined as
$g^2(0)=\langle \hat{a}^\dagger \hat{a}^\dagger \hat{a} \hat{a} \rangle / n_a^2$. 
To this end, we have numerically solved the master equation, Eq.~(\ref{equ:evol}), for the 
model hamiltonian in Eq.~(\ref{eq:model}) and the Lindblad operators in 
Eqs.~(\ref{damp}), (\ref{deph}), and (\ref{equ:px}). 
The operators are explicitly built in matrix form on a Fock basis of occupation numbers
for the atom and the cavity mode, respectively. For any given set of model parameters,
the steady state density matrix can be obtained by numerically searching for 
the eigenvector $|\rho\rangle\rangle_{ss}$ corresponding to the eigenvalue 
$\lambda_{ss}=0$ of the linear operator equation 
$\hat{L}|\rho\rangle\rangle=\lambda|\rho\rangle\rangle$. In the latter,  
$|\rho\rangle\rangle$ is essentially the density operator mapped into vectorial 
form, and $\hat{L}$ is the linear matrix corresponding to the Liouvillian operator in the 
right-hand side of the master equation. If it exists, as it is always the case for the parameters
considered, the steady state solution is unique \cite{stenholm03pra}.
After recasting the vector $|\rho\rangle\rangle_{ss}$ in matrix form, the relevant 
observable quantities can be calculated as  $\langle O \rangle_{ss} = Tr\{ \hat{O}\rho_{ss}\}$. 
In the following simulations, we kept  up to 30 photons in the basis, which is largely sufficient 
for convergence. 
We show the results in Figs.~\ref{fig2}a,b,c respectively, for $\delta=0$. First we focus on the 
case where pure dephasing is negligible (blue solid line). 
With the set of parameters used, this corresponds to the good cavity regime. Note again that 
the chosen parameters are within reach of current technology regarding quantum dots coupled 
to optical semi-conducting microcavities.
Typical coupling strengths of $g=50-100$ $\mu$eV can be reached \cite{Hennessy,Nomura}, 
whereas quality factors exceeding $Q_{cav}=10^5$ (i.e. cavity linewidths smaller than 
$\kappa = 10$ $\mu$eV) have separately been demonstrated \cite{derossi08apl}. 
However, the physics of  single two-level atom lasers and single quantum dot lasers are 
drastically different from one another. We will come back to discuss this point in Sec~D.

As it can be seen from the plotted quantities,  the device perfectly converts the pump energy
into cavity photons, whatever the pump power (log-log scale), which was expected since the 
device shows a high $\beta$ \cite{Rice,Bjork}.
Very intuitively, the critical value $n_a=1$ is reached as soon as the pump rate is of the same 
order of magnitude as the cavity damping rate. 
At this point, the atomic population remains clamped at a value nearly equal to $n_x \sim 0.5$, 
which is already a signature of lasing. This is confirmed studying the statistics of 
the emitted field, that clearly shows a transition from antibunched ($g^2(0)<1$) to 
Poissonian ($g^2(0)=1$), for nearly the same value of the pump power. Qualitatively 
similar results were recently shown in \cite{Delvalle,Elena}. 
Indeed, in the spontaneous emission regime the device produces streams of single photons, 
and the emitted field is antibunched \cite{note:antibunching}. 
When stimulated emission is reached, more than one photon can be stored in the cavity mode 
before the intra-cavity field is dissipated, leading to the buildup of a Poissonian field that 
reflects the statistics of the single atom excitation events during a typical cavity lifetime. 
Thus, in the single atom device, in addition to be an efficient relaxation channel (just as in the 
conventional laser case), the cavity plays the role of a photon delayer, which keeps the 
photons emitted by a single atom for a sufficiently long time so that a Poissonian field can build 
up in the mode. 
This crossover in the statistics of the emitted field is a signature of the transition from the 
``single photon source'' to the ``single two-level emitter laser'' operating regime.
We mention here that this behavior is quite different from what 
happens for ``conventional'' high-$\beta$ lasers involving several emitters. In the latter case indeed, 
the statistics of the field maps the statistics of the pump, whatever its power is \cite{Rice}. 
The single photon source regime has been observed, e.g., for a Caesium atom strongly 
coupled to an optical cavity \cite{MacKeever}. 
A crossover to Poissonian statistics has been observed - to a certain extent -
for a single quantum dot coupled to a micropillar cavity \cite{Reitzenstein}, and a photonic 
crystal cavity \cite{Nomura,Nomura2}.   
The single quantum dot laser case, and its differences and similarities with respect to the two level atom case, 
are discussed in Subsec. D.

When the pumping rate is too large, the atomic emission becomes incoherent, leading to 
a decrease of the cavity population, and to the corresponding increase of the atomic 
population, until total inversion in reached. The emitted field becomes thermal and the 
parameter $g^2(0)$ converges to its limiting value, 2. This phenomenon is known as the 
quenching of the laser, and was first predicted in \cite{Mu}; it was attributed to the saturation 
of the two-level emitter. The notion of good cavity regime sheds new light on this feature.
As a matter of fact, the atom-cavity coupling $R$ decreases with respect to pump power 
$P_x$. Once the lasing regime is reached, one can increase $P_x$, and consequently 
the number of photons in the lasing mode, as long as one remains in the good cavity regime; 
when this condition is no longer satisfied, the laser gradually switches off. 
Thus, in this view quenching is due to the transition to the bad cavity regime, induced by 
power broadening. 
We have plotted the evolution of the parameter $R$ with respect to pump power $P_x$ 
(inset): as it can be seen in the figure, quenching happens for a typical value of the pump 
$P_x \sim 20g$, for which the atom-cavity coupling constant becomes lower than $\kappa$, 
confirming our initial guess.

Before examining the influence of pure dephasing on the lasing signatures, we mention 
another usual criterium for lasing in conventional devices, namely the narrowing of the 
spectrum emitted by the cavity mode. In the single atom device, it has been predicted 
by \cite{Walther,Delvalle,Elena}, and maybe experimentally observed in \cite{Nomura2}, 
that at low pump power the spectrum consists in a series of peaks (the so-called  
``Jaynes-Cummings forks''  \cite{Delvalle}).  The authors of \cite{Nomura2} 
have talked about a ``coexistence of the strong coupling regime and the lasing regime''. 
We underline here that the emission  of Jaynes-Cummings forks is a natural feature of the 
single emitter laser \cite{Walther}. 
Reaching stimulated emission in the steady state regime simply means that one observes 
photons coming from radiative transitions between states of the excited manifolds in the 
Jaynes Cummings ladder,  which naturally results in series of peaks centered around the 
cavity frequency. 
Increasing the pump power leads to the broadening of the peaks and to their convergence 
towards a single one at the cavity frequency, as mentioned in \cite{Walther,Delvalle,Elena}. 
The transition between multi-peaks and single peak emission takes place after the lasing 
threshold, as evidenced in \cite{Nomura2}.

\subsection{Influence of detuning and pure dephasing}

\begin{figure}[t]
\begin{center}
\includegraphics[width=9cm]{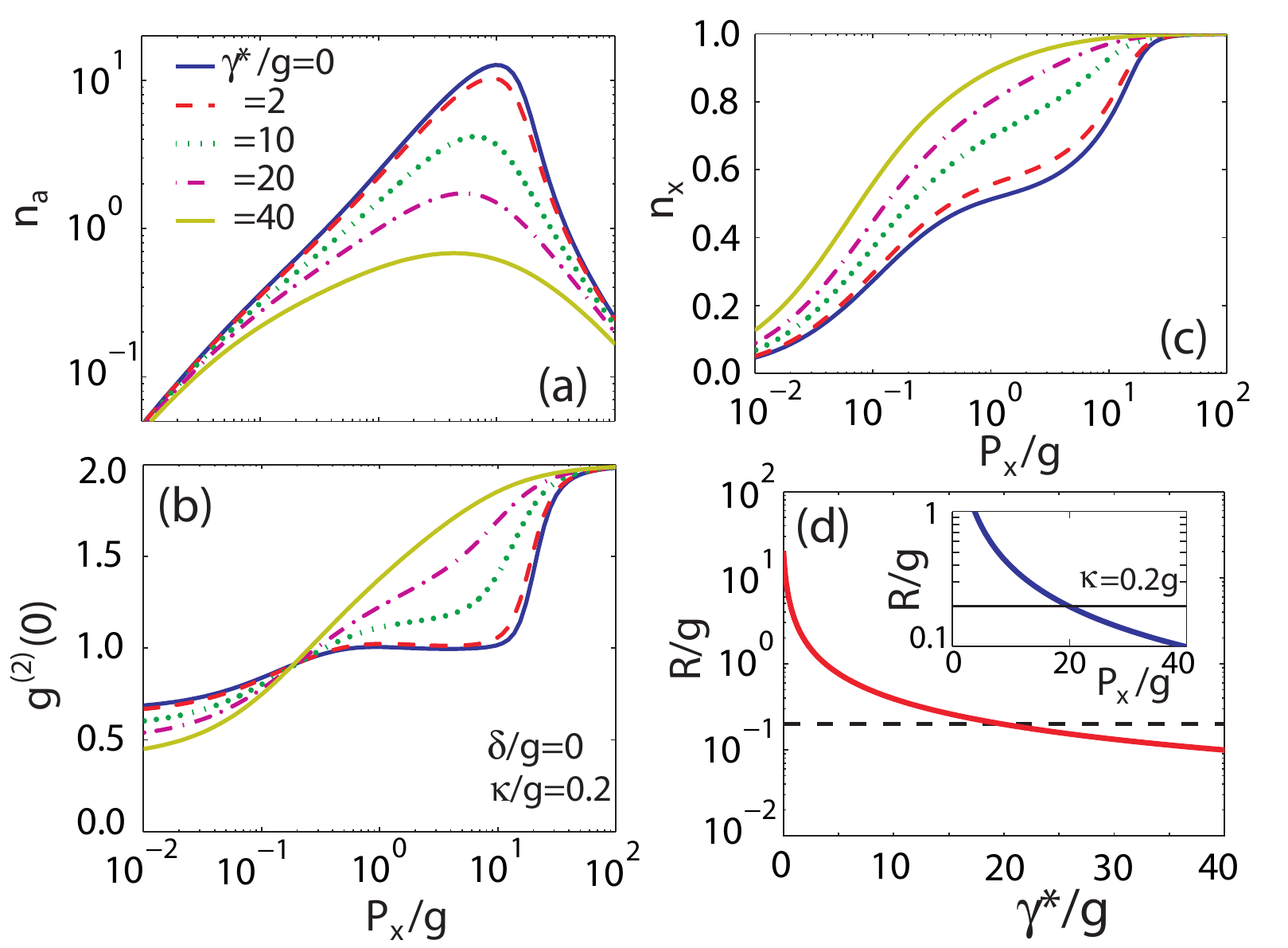}
\caption{(Color online).
Numerically calculated (a) cavity population, 
(b) statistics of the emitted field, and (c) atomic population as 
a function of pump power, for different pure dephasing rates. 
(d) Effective atom-cavity coupling as a function of the pure dephasing rate (solid line), 
as compared to the cavity damping rate (dashed line).  Parameters are: 
$\kappa=0.2g$ and $\delta=0$. The inset shows the effective atom-cavity coupling as a 
function of $P_x$, for $\delta=0$, $\gamma^{\ast}=0$.} \label{fig2}
\end{center}
\end{figure}

\begin{figure}[t]
\begin{center}
\includegraphics[width=9cm]{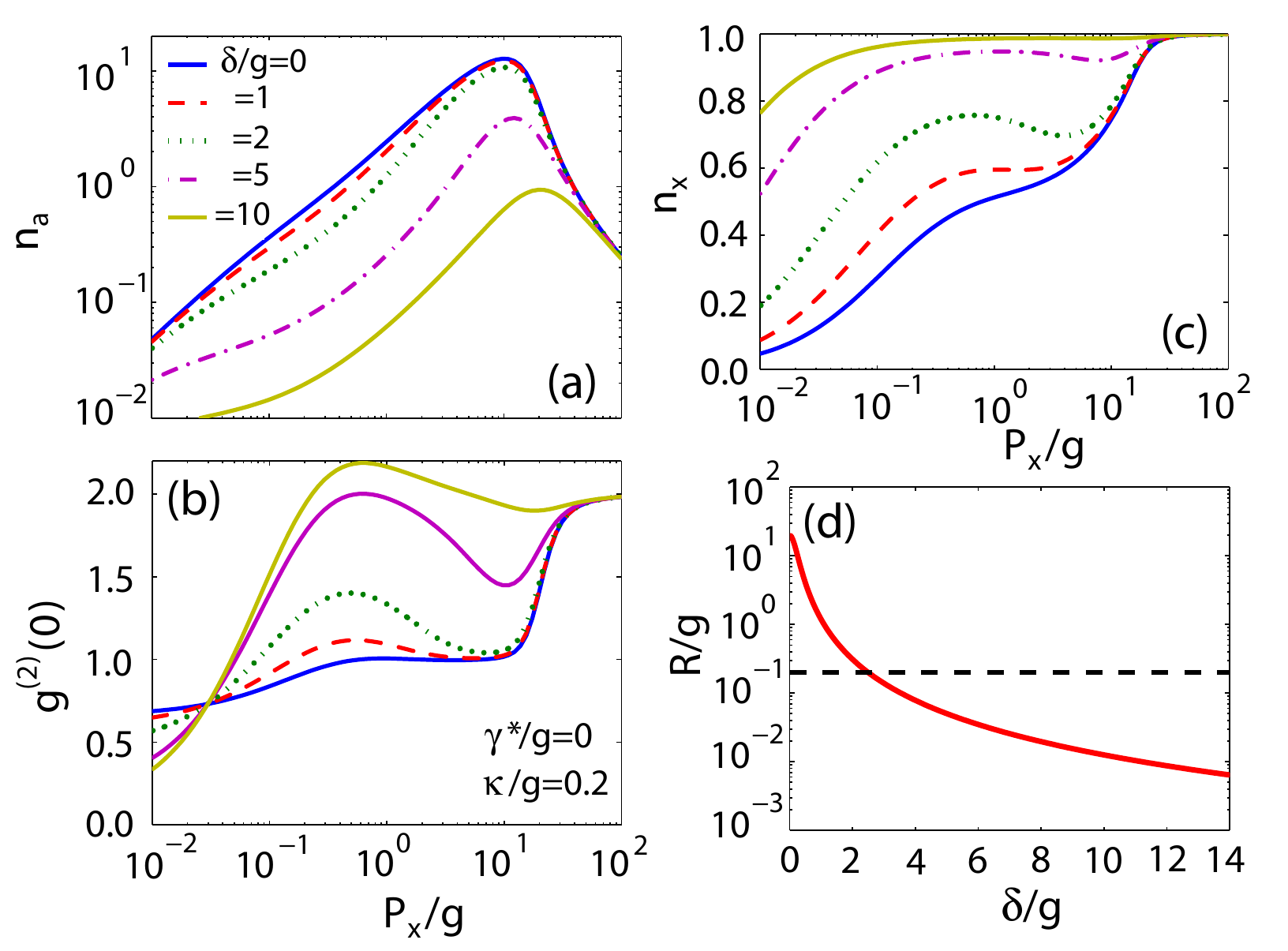}
\caption{(Color online).
Numerically calculated (a) cavity population, (b) statistics of the emitted field as 
a function of pump power, and (c) atomic population for different atom-cavity detunings. 
(d) Effective atom-cavity coupling as a function of the detuning (solid line), 
as compared to the cavity damping rate (dashed line).  Parameters are:
$\kappa=0.2g$ and $\gamma^{\ast}=0$.} \label{fig1}
\end{center}
\end{figure}

In Subsec.~A, we have evidenced that a necessary condition for lasing is to achieve the 
good cavity regime. In the previous Sections we have seen that pure dephasing strongly 
influences the effective atom-cavity coupling rate. As a consequence, one expects that it 
should affect the lasing conditions of the system as well.
When considering the resonant case of Fig.~\ref{fig2}, we see that increasing 
$\gamma^{\ast}$ lowers the effective coupling $R(\delta, \gamma^{\ast})$, up to 
the point where the lasing criteria are completely lost (e.g., curves for $\gamma^{\ast} =40g$ 
in Figs.~\ref{fig2}a and b). 
In particular, the clamping of the autocorrelation function to the Poissonian value $g^2(0)=1$ 
disappears, and the emitted field continuously evolves from antibunched to thermal without 
showing any coherent character. 
Loss of the lasing criteria appears for a value of the pure dephasing rate 
$\gamma^{\ast} \sim 20g$. 
As evidenced in Fig.\ref{fig2}c, this corresponds to the transition from the good 
to the bad cavity regime, confirming our previous intuition.
By the very same mechanism, lasing can also be lost by increasing the atom-cavity 
detuning, as it appears by plotting the same quantities in Fig.~\ref{fig1}. 
Starting from $\delta=0$, the switching from good to bad cavity regime happens for 
$\delta \sim 2g$, which again yields the disappearance of any lasing signature. 

\begin{figure}[t]
\begin{center}
\includegraphics[width=9cm]{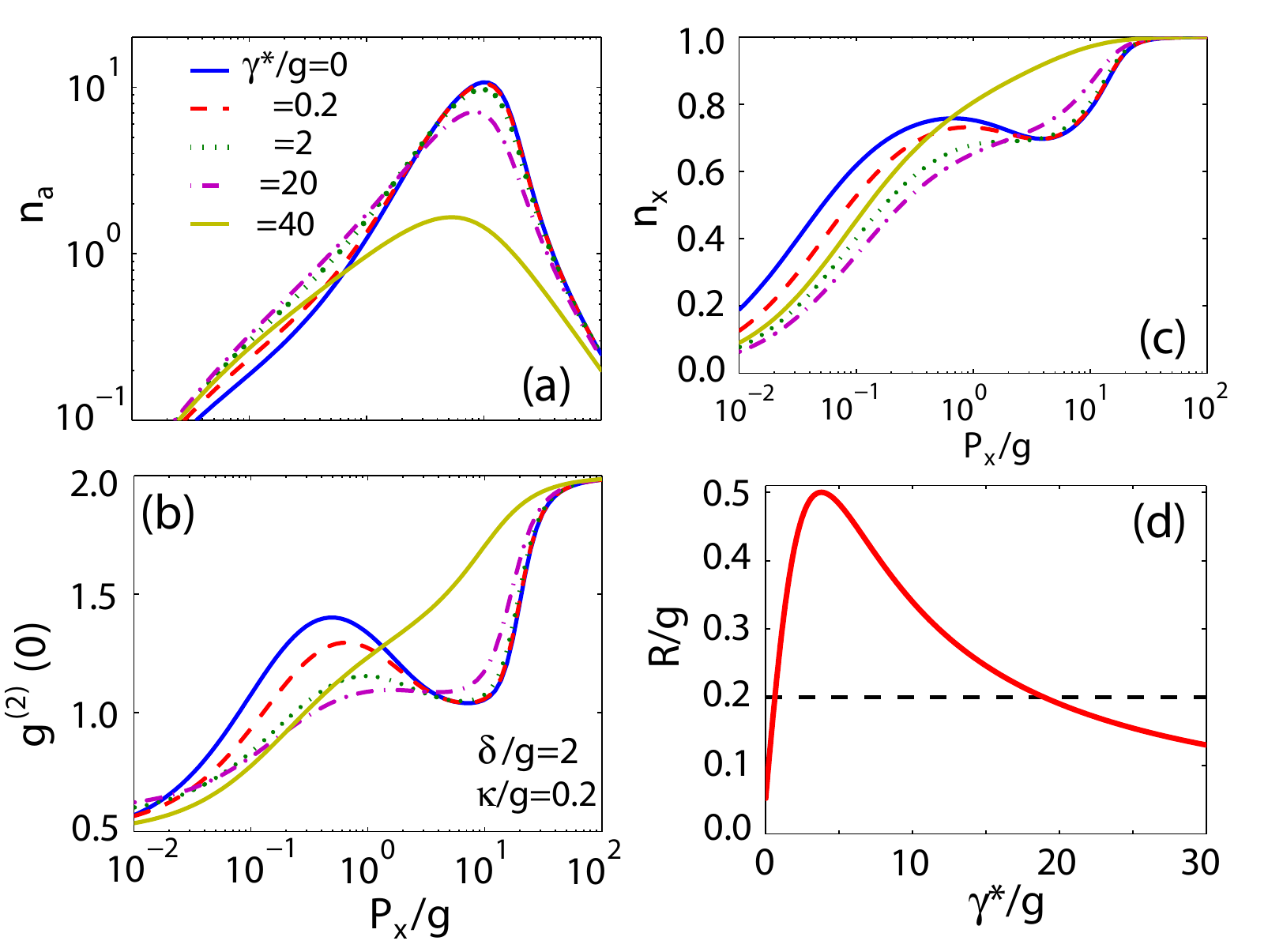}
\caption{ (Color online)
Numerically calculated (a) cavity population, (b) statistics of the emitted field as 
a function of pump power, (c) atomic population, for different pure dephasing rates. 
(d) Effective atom-cavity coupling as a function of the pure dephasing rate (solid line), 
as compared to the cavity damping rate (dashed line).  Parameters are: 
$\kappa=0.2g$ and $\delta=2g$.} \label{fig3}
\end{center}
\end{figure}

On the other hand, we have seen in Sec.~II that if the atom and the cavity are detuned, 
increasing pure dephasing can even increase the effective coupling between the two systems. 
This induces a transition from the bad cavity to the good cavity 
regime, and it allows to recover the lasing conditions. 
This result is shown in Fig.~\ref{fig3}c for a typical pure dephasing rate 
$\gamma^{\ast}\sim 2g$; 
in particular, one recovers a clear clamping of the autocorrelation function to $g^2(0)\sim 1$,
as in Fig.~\ref{fig2}b for $\delta=0$ and $\gamma^{\ast} =0$.
In other words, under such conditions it turns out that pure dephasing compensates for 
atom-cavity detuning. 
This effect, which in the previous Section was responsible for an improvement of the single
photon source figure of merit (for system parameters in the bad cavity regime), 
leads here to a recovery of the lasing signatures. 
Again, pure dephasing appears as a valuable resource for solid-state nanophotonic devices.

%
Finally, we discuss now the interest of pure dephasing in the frame of conventional lasers.
If one is just interested in efficient energy conversion, lasers involving a high number of
emitters are naturally to be preferred over the single emitter device, as they are less subject to 
saturation and quenching. Nevertheless, the criterium of high $\beta$ is challenging to 
realize for each emitter, because of inhomogeneous broadening in the solid-state 
environment, or the atomic motion in gas lasers.
In such cases, pure dephasing could provide an effective tool to overcome this problem. 
As a matter of fact, increasing the homogenous linewidth of a bunch of detuned emitters 
would not only help reaching the good cavity regime, which is less critical to fulfill in the 
N emitter's case, but would also increase their individual $\beta$ factors so that low-threshold
lasing could be favourably achieved. We stress that pure dephasing here is nothing but a very 
effective model for the broadening of solid-state emitters because of their interaction with 
the solid-state matrix. Still, such a mechanism may explain some unconventional lasing 
characteristics of few quantum dots detuned from high-quality photonic crystal cavity modes 
\cite{Strauf}, and again, it looks promising in the context of high $\beta$ multi-emitters lasers.

\subsection{Single quantum dot lasers}

The potential of a single QD coupled to a semiconducting cavity to achieve a solid-state 
single emitter laser has been explored theoretically \cite{Pelton,Benson,Delvalle,Elena},
and recent experimental demonstrations tend to show that laser gain at the single quantum 
dot level is within reach \cite{Reitzenstein, Nomura, Nomura2}.
However, as it was underlined by the authors themselves \cite{Nomura,Nomura2}, a 
laser based on single QD emission does not simply maps the physics of the single 
atom-laser into a solid-state system. 
One main source of differences is that solid-state cavities are always coupled to a bunch 
of background emitters that can efficiently feed the mode even if the emitters and the cavity 
are detuned, as it was experimentally evidenced in \cite{Hennessy, Press, Kaniber, Pascale}, 
and theoretically explored in \cite{Winger, Naesby, Alexia}. Cavity feeding is generically
attributed to the Purcell enhancement of relaxation processes that are resonant with the 
cavity mode, the very existence of these processes being due to phonon-assisted 
decay \cite{Winger} or pure dephasing \cite{Alexia, Naesby} that broaden the 
emitters' linewidths. 
Because of this background, cavity field at low pumping rate is not totally antibunched,  as 
it also appears in \cite{Reitzenstein, Nomura, Nomura2}, giving a quantitative measurement 
of the contribution of the single QD to the cavity emission \cite{note:antibunching}. 

Most importantly, even in the ideal situation where only a single QD is coupled to the cavity 
mode, the usually employed non-resonant pumping scheme to operate lasers allows to 
pump more than one exciton in the QD, making the physics of the single QD laser  essentially 
different from a single atom laser.  
As multiexcitonic transitions happen at different  frequencies because of 
exciton-exciton interaction, it was initially thought that such a  device would suffer from 
blinking \cite{jmg}. In fact, it can be intuitively argued that if the excitonic transition of the QD 
is resonant  with the cavity mode, the QD decouples from the cavity as soon as it contains 
two  excitons, thus leading to the device switch off.

Although, experimental evidence has shown that single QDs-based lasers do not suffer
from blinking, and their lasing transitions display different statistical behaviors ranging 
from  partial antibunching to Poissonian \cite{Nomura,Nomura2}.  
In this context, the evolution of the statistics of the cavity light field with respect to the pump 
power has not the same physical meaning as in the two-level atom case, which has been 
the subject of the present work,  as it was also underlined in \cite{Nomura2}. 
As a matter of fact, because of pure dephasing, phonon-assisted processes, or power 
broadening  (as it was evidenced above),  increasing pump power leads to a broadening 
of the higher order excitonic transitions,  that can also contribute to cavity feeding. 
In the limiting case where all the transitions are efficiently coupled to the cavity, 
the statistics of the cavity field will simply map the Poissonian statistics related to the 
number of excitons in the QD. 
As a consequence, the role played by the cavity is drastically different from the single atom 
case:  it does not act as a photon trap, delaying the emission of the field until a Poissonian 
statistics  has built in the mode, but rather as a common relaxation channel for all the 
transitions of the QD. Thus, the crossover from the antibunched to the Poissonian 
field just reflects the excitation and the broadening of the multiexcitonic transitions.
Note that the observation of such a transition does not require the good cavity regime, 
which explains why lasing was also observed in Refs. \cite{Reitzenstein,Nomura}, 
where  the QD and the cavity mode were only weakly coupled. 

\begin{figure}[t]
\begin{center}
\includegraphics[width=9.5cm]{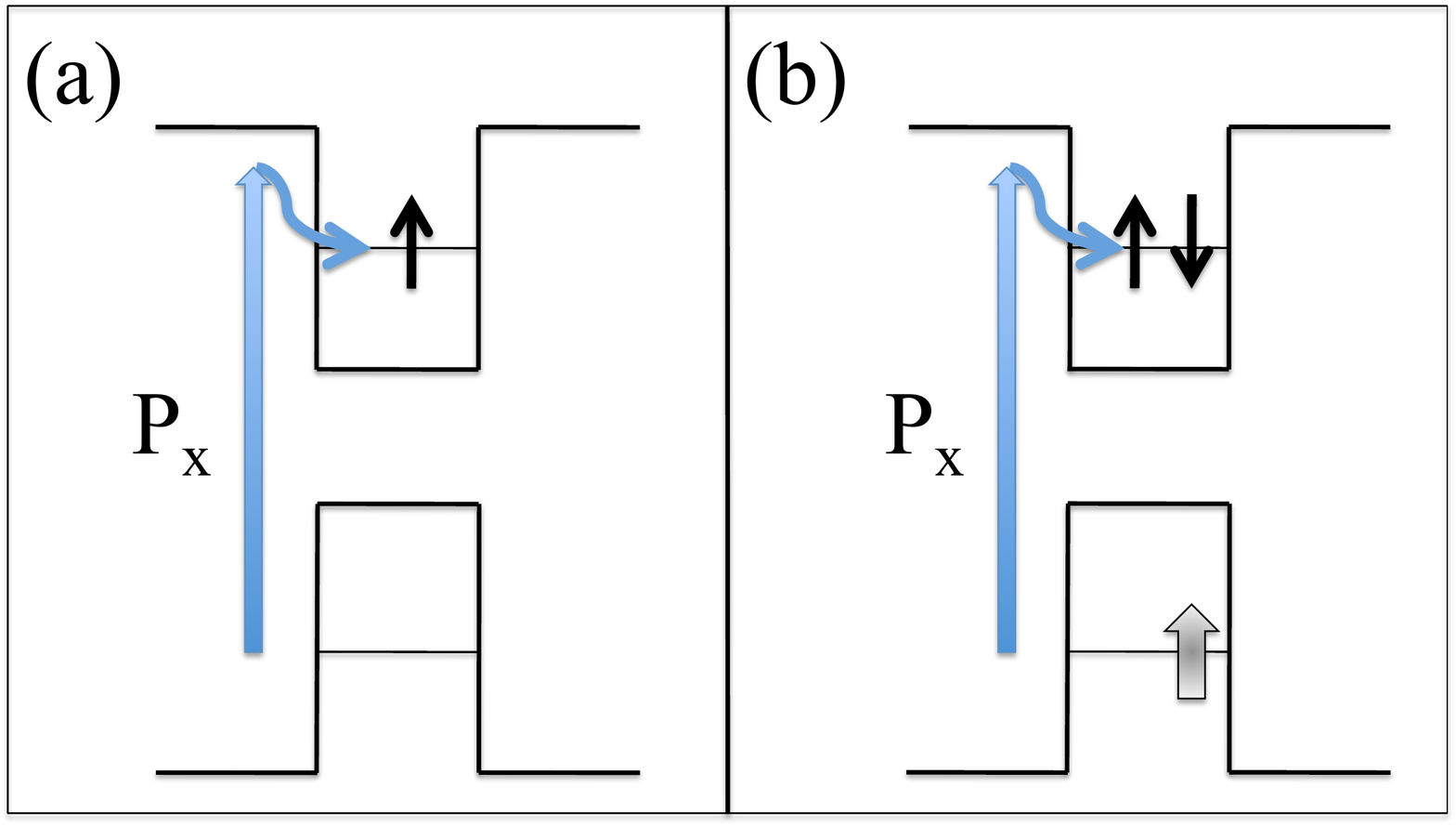}
\caption{ Proposal to achieve a single atom-like laser using a QD doped with a single electron. 
A quasi-resonant pumping of the QD using a phonon-assisted optical transition 
is used to excite a trion. Due to this particular pumping protocol, the emitter has only two 
possible states: (a) no exciton in the QD;  (b) a single exciton in the QD.} \label{fig:SQDL}
\end{center}
\end{figure}

We propose an alternative strategy to realize a single atom-like laser with a QD coupled 
to a  cavity mode, namely, to use a QD doped with a single electron embedded in a 
high-$Q$/low-$V$  microcavity in the good cavity regime  (see schematic picture  in 
Fig.~\ref{fig:SQDL}). Such a device, including tuning parameters to control  the charge state 
of  the QD and the QD-cavity detuning, is within reach of current technology 
\cite{Strauf2, Finley2}.  One can selectively inject an additional exciton in the QD through 
a quasi-resonant optical pumping, using a transition assisted by the creation of a phonon 
(either optical or acoustic). If the energy levels in the QD are energetically well separated, 
the injection of an additional exciton is forbidden because of Pauli blocking. This experimental 
approach will permit to realize the device proposed in the present  manuscript, namely 
a compact and solid-state based system that can be operated  either as a single photon 
source or as a nanolaser, depending on the pumping and pure dephasing rates. 
In particular, the dramatic influence of  pure dephasing on the lasing threshold could be 
evidenced on such a device. 

\section{Conclusion}

We have studied the dynamics of a two-level atom undergoing pure dephasing, 
coupled to a single cavity mode,  and derived an effective atom-cavity coupling rate, 
which has been shown to be a useful and conceptually simple parameter to be used 
in the description of the physics of the problem. In particular, it  allowed us to generalize 
the notions  of good and bad cavity regimes. 
In the bad cavity regime, we have defined a generalized Purcell factor, and studied 
the strategies to optimize broad emitter-based single photon sources. We have shown 
that if the atom and the cavity are detuned, increasing pure dephasing can improve the
figures of merit of the device. 
In the same way, we have shown that in the good cavity regime, lasing can  even be 
induced by increasing pure dephasing. 
These results  enforce the idea that pure dephasing is a promising resource, specific 
to solid-state emitters, which might be used to develop advanced nano-photonic devices
like single photon sources and nanolasers, and could allow to improve 
their  performances. These ideas could be directly checked on an innovative
and versatile device based on the control by pure dephasing of the coupling between
a single QD and a semi-conducting cavity. 

\begin{acknowledgments}
This work was supported by NanoSci-ERA consortium and by the EU under 
ERANET project LECSIN, by the Nanosciences Foundation of Grenoble, 
the CNPq, the Fapemig, and the ANR project CAFE. 
DG acknowledges stimulating discussions with A. Imamo\v{g}lu and T. Volz.
\end{acknowledgments}

\newpage

%
%

\begin{thebibliography}{99}
\bibitem{note:antibunching}
We notice that even in the absence of dephasing or other dissipation channels, 
the antibunching is never perfect, due to the presence of the interplay between 
incoherent pumping and cavity losses ($P_x$ vs. $\kappa$).
\end{thebibliography}
\end{document}